\documentclass[amssymb,amsmath,prl,amsfonts,superscriptaddress,twocolumn]{revtex4}

\usepackage{graphicx}
\usepackage{color} 
\definecolor{bleu}{rgb}{0,0,0}

\usepackage{bm} 
\usepackage{hyperref} 
\begin{document}

\title{Distortion and destruction of colloidal flocks in disordered environments}
\author{Alexandre Morin}
\author{Nicolas Desreumaux}
\author{Jean-Baptiste Caussin}
\author{Denis Bartolo}
\affiliation{Univ. Lyon, Ens de Lyon, Univ. Claude Bernard, CNRS, 
Laboratoire de Physique, F-69342 Lyon, France}
\maketitle

\textbf {How do flocks, herds and swarms proceed through disordered environments? This question is not only  crucial to  animal groups in the wild, but also to virtually  all applications of collective robotics, and  active materials composed of  synthetic motile units~\cite{Handbook,Nagpal2014,Berman,Shklarsh,Baush2010,Deseigne,Thutupalli,Dogic,Bocquet,Bricard2013,Palacci,Bechinger,Goldstein,Sano2015}. 
In stark contrast, appart from very rare exceptions~\cite{Peruani1,Olson2014,Quint}, our physical understanding of flocking  has been hitherto limited to 
homogeneous media~\cite{Marchetti_review,Vicsek_review,Cavagna_review}. Here we explain how collective motion  survives to geometrical disorder. To do so, we combine experiments  on  motile colloids  cruising through random  microfabricated obstacles, and analytical theory. We explain how disorder and bending elasticity compete to channel the flow of  polar flocks  along sparse river networks akin those found beyond plastic depinning  in driven condensed matter~\cite{Reichhardt_Review}.
Further increasing disorder, we  demonstrate that   collective motion is suppressed in the form of  a  first-order phase transition generic to  all polar active materials. %
}

\begin{figure*}
\begin{center}
\includegraphics[width=\columnwidth]{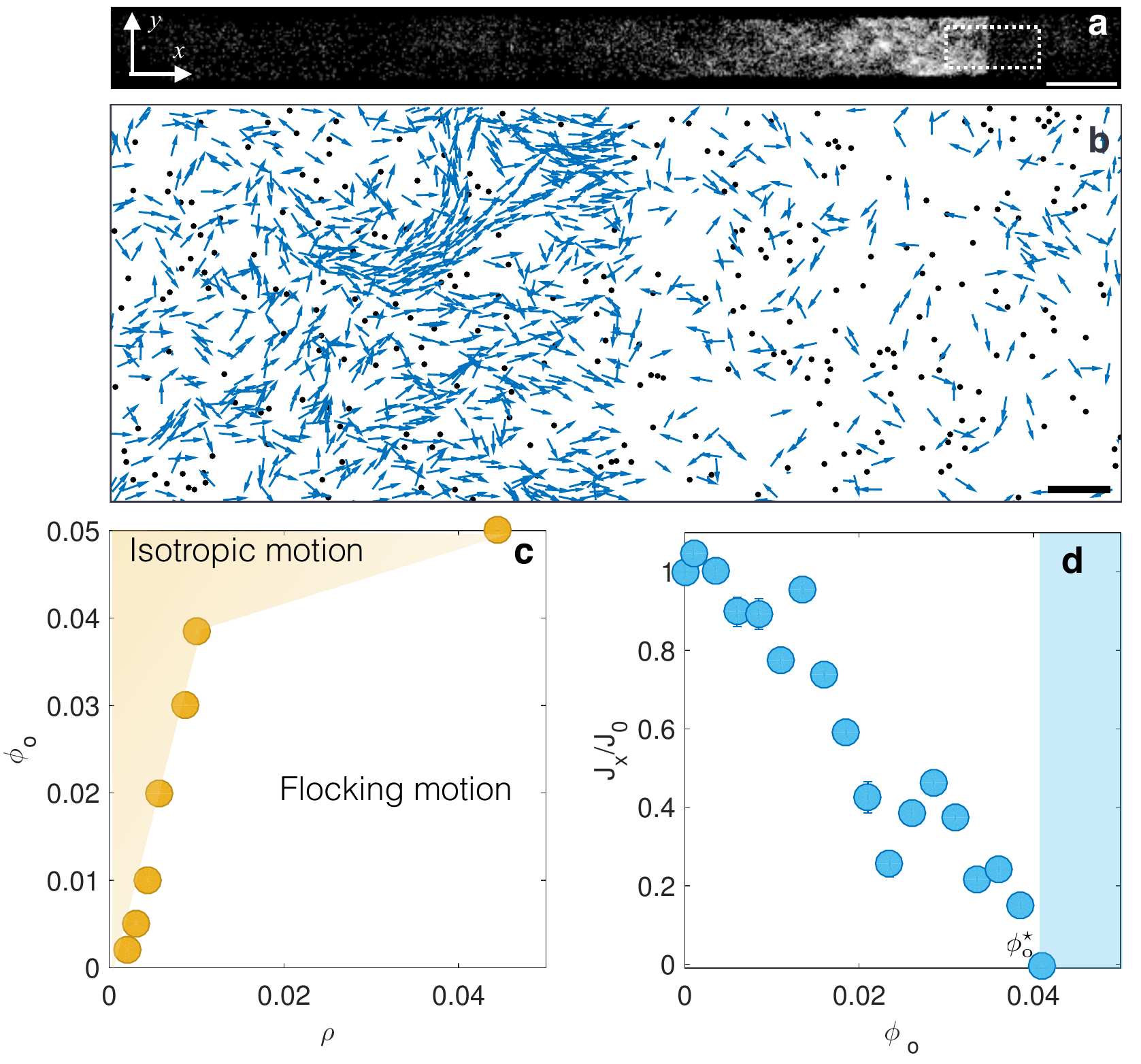}
\caption{{\bf Emergence and suppression of collective motion.}
\textbf a, Stitched fluorescent images of a 7 mm-long colloidal flock cruising in a rectangular channel. Total number of colloids: 8,500. Dotted rectangle:  region in which the velocity measurements of panel
 \textbf b are performed. Scale bar: 1 mm. \textbf b, Close-up on the head of a colloidal swarm propagating past random obstacles (black dots). The arrows are located at the colloid positions and point along the orientation of their velocity. Obstacle packing fraction: $\phi_{\rm o}=2.45\times10^{-2}$. Scale bar: $100\,\rm \mu m$. 
\textbf c, Flocking phase diagram in the $(\rho,\phi_{\rm o})$ plane. The symbols represent the variations of $\phi_{\rm o}^\star$ with $\rho$. Error bars: smaller than the symbols (Defined as the difference between the minimal  value of $\rho$  above  which flocking was observed and the maximal value below which isotropic motion only was observed). 
\textbf d, The $x$-component of the roller current  is normalized by $J_0$  measured in an obstacle-free channel. $J_x/J_0$ is plotted as a function of the fraction of obstacles. Orientational order is  suppressed in the shaded region. Error bars: 1 sd (17 different flocks).
\label{fig1}
}

\end{center}
\end{figure*}

We use the  experimental system introduced in~\cite{Bricard2013,Bricard2015}, which  consists in colloidal rollers powered by the so-called Quincke electro-rotation mechanism~\cite{Quincke}, see Methods and Supplementary Methods.  
The motile colloids experience both hydrodynamic and electrostatic interactions  which  promote  alignement of their translational velocity~\cite{Bricard2013,Bricard2015}. 
When the roller packing fraction, $\rho$, exceeds $3\times10^{-3}$, these  polar interactions overcome  rotational diffusion and macroscopic collective motion emerges~\cite{Bricard2013,Bricard2015}. In the homogeneous slab geometry shown in Fig.~\ref{fig1}a, a seven-millimeter-long  flock  spontaneously forms and cruises through a dilute ensemble of rollers moving isotropically, see Supplementary Video 1. The flock has a sharp front, a long tail,  and endlessly cruises at a constant speed  along the $x$-axis, bouncing back and forth on the confining walls.  The flock speed $c_{\rm F}$ is found to be equal  to the speed of an isolated roller $v_0=1.4\pm0.1\,{\rm mm\ s^{-1}}$.  

Can flocks propagate in disorder media? How does this broken-symmetry phase survive to geometrical disorder? 
In order to answer these questions, we include  randomly distributed circular obstacles of radius $a=5\,\mu\rm m$ in the microfluidic channel. When the obstacle packing fraction $\phi_{\rm o}$ is small, collective motion still emerges according to the same nucleation and propagation scenario, see Fig.~\ref{fig1}b and Supplementary Video 2.  However as $\phi_{\rm o}$ exceeds a critical value, $\phi_{\rm o}^\star$, the  obstacle collisions suppress  any form of global orientational order and macroscopic transport. Correlated motion persists only at short scales, as illustrated in Supplementary Video 3. 
As expected, dense flocks are more robust to disorder and $\phi_{\rm o}^\star$ monotonically increases  with the roller fraction $\rho$,  Fig.~\ref{fig1}c.

In all that follows,  the sole control parameter of our experiments is the obstacle fraction $\phi_{\rm o}$. The roller fraction  is set to a constant value  above the flocking threshold in a obstacle-free channel, $\rho=(1.02\pm0.06)\times 10^{-2}$. 
A natural order parameter for the flocking transition is  the magnitude $J_x$  of the  roller current $\textbf J(\textbf r,t)$ projected on the $x$-axis, and averaged over  time and space. Accordingly $J_x$ monotonically decreases with $\phi_{\rm o}$ and vanishes at $\phi^\star_{\rm o}$, Fig.~\ref{fig1}d. 

\begin{figure*}
\begin{center}
\includegraphics[width=\columnwidth]{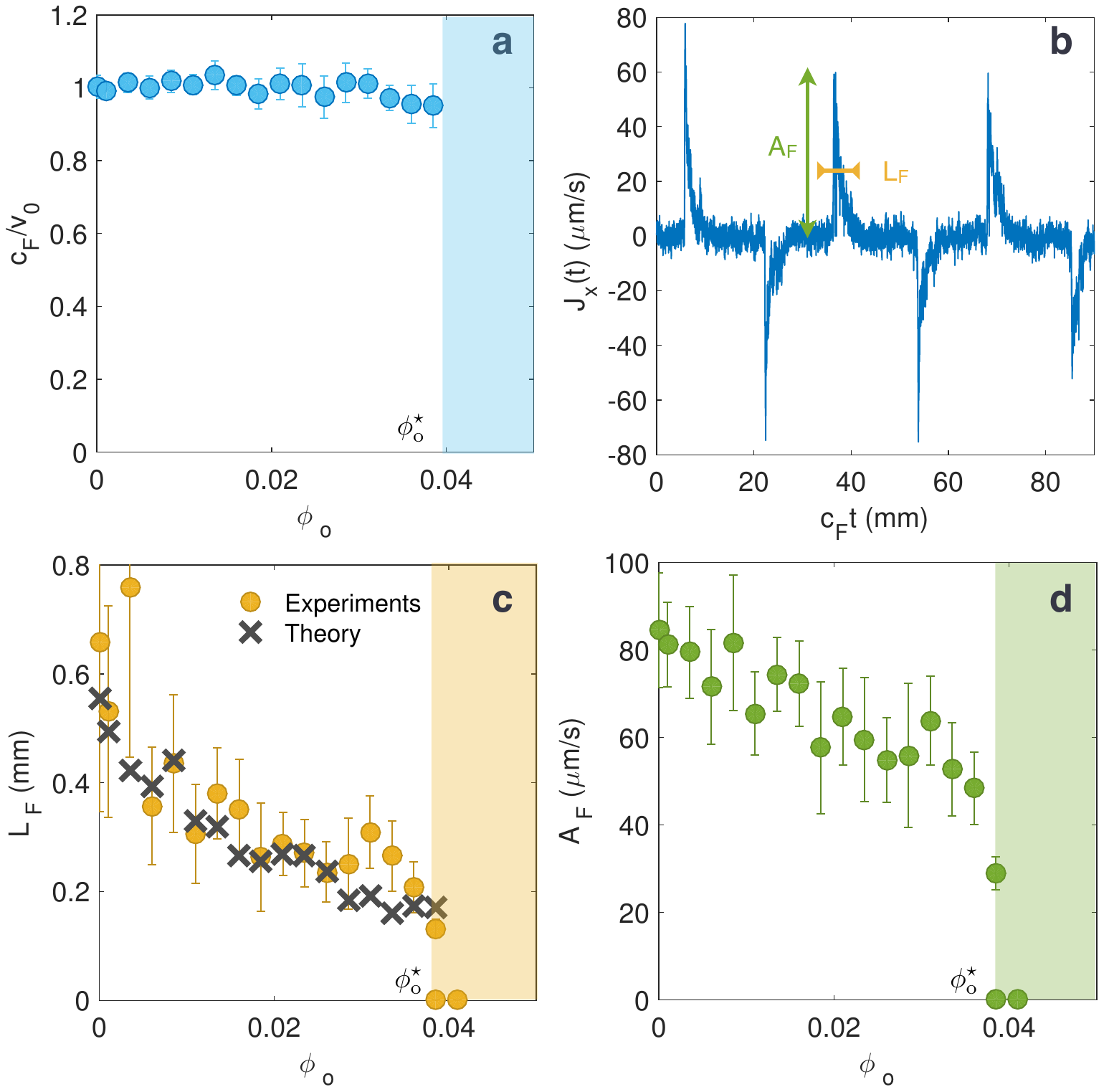}
\caption{{\bf Flock morphology.}
 \textbf a, Flock speed normalized by the  roller speed plotted as function of the obstacle fraction. Error bars: 1 sd (17 flocks per data point). 
 \textbf b, Spatial variations of the longitudinal current. The shape of the steadily propagating flock is readily inferred from the temporal variations of the current averaged in a $25\,\rm\mu m\times1mm$  rectangular region. The flock width $L_{\rm F}$ is defined as the width of $J(t)$ at $2/3$ of the maximal amplitude $A_{\rm F}$.
 \textbf c, Yellow symbols: Flock length plotted versus the obstacle fraction (averaged over 17 different flocks).  Error bars:  1 sd. Black symbols: Analytical prediction of the flock length, see Supplementary Note. 
 Shaded region: isotropic phase.  
 \textbf d, Maximal amplitude of the longitudinal current plotted versus the obstacle fraction (averaged over 17 different flocks). Shaded region: isotropic  phase.  Note that at $\phi_{\rm o}^\star$ both a flock state and a homogeneous isotropic state  coexist. Error bars:  1 sd.}
\label{fig2} 
\end{center}
\end{figure*}

Our  first goal is to elucidate this loss of orientational order. To do so, we  consider the evolution of the flock  morphology along the propagation direction upon increasing disorder.  This morphology can be equivalently captured by the variations of the local density, current, or polarization  fields as demonstrated in a Supplementary Note. For sake of clarity we focus here on the roller current as the main observable.
The flock speed $c_{\rm F}$ is unaltered by disorder and remains very close to the roller velocity for all $\phi_o<\phi_{\rm o}^\star$, Fig.~\ref{fig2}a. Therefore  the time variations of $J_x(t)\equiv\langle \hat{\mathbf x}\cdot\mathbf J(x=0,y,t)\rangle_{y}$,  the longitudinal current  averaged over the transverse direction, give an accurate description of the coarse-grained shape of the flocks, see Fig.~\ref{fig2}b. 
Three important results are in order: the decrease of the flock length, $L_{\rm F}$, echoes that of the global order parameter and  vanishes  rather smoothly at $\phi^\star_{\rm o}$, Fig.~\ref{fig2}c.  However, as shown in Fig.~\ref{fig2}d, the maximal current amplitude,  $A_{\rm F}$,  undergoes a sharp drop and cancels discontinuously  at $\phi^{\star}_{\rm o}$. Finally, at $\phi^{\star}_{\rm o}$,  the flocks are intermittent:  they repeatedly form and propagate steadily, before spontaneously vanishing and nucleating again. A  featureless isotropic state coexists in time with a phase-separated flocking state where macroscopic  excitations as large as $1\,\rm mm$  propagate in the channel. Altogether these three observations  firmly evidence that  disorder suppresses  the flocking state in the form of a  first-order  non-equilibrium  transition.
 
However, the obstacles do not merely reduce the extent of the flocks down to their extinction but also trigger qualitative changes in their inner structure. {\color{bleu}The  snapshots of the  roller current $\mathbf J(\mathbf r,t)$ at four subsequent times in Fig.~\ref{fig3}a  demonstrate that the flocks are strongly heterogeneous spatial patterns, see also Supplementary Video 4. We characterize the local flock morphology by introducing the current field $ \bm{J}_{\rm flock}(\mathbf r)=\langle \mathbf J(\mathbf r,t)\rangle_{t\in\Delta t_{\rm F}}$  averaged over the time interval $\Delta t_{\rm F}$ taken by the  flock to cross the observation window.} At low $\phi_{\rm o}$, we observe that colloid-depleted wakes  as large as $\sim50\,\rm\mu m$ form downstream each  obstacle, see Fig.~\ref{fig3}b upper panel. {\color{bleu}However, as $\phi_{\rm o}$ increases, the competition between alignment interactions  and  multiple-obstacle scattering, causes the redistribution of the roller current into a  {\em static} river network, Fig.~\ref{fig3}b.  Virtually no collective motion  occurs   in the closed regions surrounded by the flowing rivers (black regions in Fig.~\ref{fig3}b). The  extent of the  regions where the flow is suppressed   can significantly exceed both  the typical inter obstacle-distance and the depletion-wake size. 
Importantly, upon increasing $\phi_{\rm o}$, the river network becomes increasingly sparse and different from  the region of space left around  the mere  superposition of uncorrelated wakes.  Comparing the areas of these two very different geometries allows us to quantify the  sparsity of the river networks  in Fig.~\ref{fig3}c.} In addition, the  networks also become increasingly tortuous as demonstrated by the fast increase of the orientational fluctuations of $ \bm{J}_{\rm flock}$ with $\phi_{\rm o}$ in Fig.~\ref{fig3}d.
{\color{bleu}Above $\phi^\star_{\rm o}$ orientational order survives to disorder in finite and short-lived rivers. Any form of macroscopic transport is suppressed as these transient channels are isotropically distributed and  do not percolate through the entire system, see Supplementary Video 3 and Supplementary Note.}
We close this discussion by  stressing that these emergent river networks are strikingly similar to that encountered above the plastic depinning threshold when driving an ensemble of elastically-coupled particles through quenched disorder (from vortex lattices in type-II superconductors, to driven colloids and grains)~\cite{Jensen88,Gronbech96,Troyanovski99,Ledoussal98,Faleski96,Topinka2001,Pertsinidis2008,Reichhardt_Review,Wyart2015}. 

In order to elucidate the physics underlying the suppression of collective motion and the emergence of channelling networks, we first need a  quantitative description of the roller-obstacle interactions. As a roller approaches an obstacle its direction of motion is repelled at a distance yet its speed remains unchanged, Figs.~\ref{fig4}a and \ref{fig4}b.  The roller-obstacle and roller-roller repulsions stem from the same physical mechanisms~\cite{Bricard2013}: a dielectric obstacle causes a local radial perturbation of  the  electric field ${\mathbf E}=E_0\hat{\mathbf z}$ used to power the Quincke rotation. As a result, a short-range repulsive {\em torque} reorients the roller velocity in the direction opposite to the obstacle, see Supplementary Note. This interaction has the same symmetry as that numerically considered in~\cite{Peruani1,Peruani2}.  The scattering plots shown in Fig.~\ref{fig4}c and Supplementary Note, demonstrate that the repulsive torques are weak and short ranged. A head-on collision merely deflects the initial roller orientation by an angle of $60^\circ$. As a consequence, up to $\phi_{\rm o}=0.1$,  the trajectories  in the isotropic phase remain diffusive at long times, and are fully characterized by their  rotational diffusivity $D$, which linearly increases  with $\phi_{\rm o}$, Fig.~\ref{fig4}d. 

{\color{bleu}We can now account for the  first-order nature of the  flocking transition. The linear increase of $D$ suggests simplifying  the  interactions between the rollers and the obstacles  as uncorrelated binary collisions  with random scatterers~\cite{Peruani2}.} Within this Boltzmann approximation, we can generalize the kinetic theory valid at the onset of collective motion, which we introduced in~\cite{Bricard2013}. We show in a Supplementary Note that the roller-obstacle and roller-roller interactions decouple. Increasing the obstacle fraction solely renormalizes the angular noise acting on the rollers, even when they  interact in the flock phase. We then readily conclude that the transition to collective motion should belong to the very same universality class as the {\em first order} flocking transition found in all motile-spin models,  starting from the seminal Vicsek model~\cite{Vicsek95,Chate2004,Solon2015}. We quantitatively test the relevance of this scenario, by comparing in Fig.~\ref{fig2}c,  the measured flock length to our theoretical prediction for the shape of such non-linear excitations as detailed in a Supplementary Note.
The  unambiguous agreement confirms  our theoretical explanation: weak quenched disorder triggers a generic Vicsek-like discontinuous transition from collective to isotropic motion. 

However this appealing scenario cannot capture the emergence of channeling networks at high obstacle fractions. {\color{bleu}We now theoretically account for these spatial fluctuations by describing  the strongly polarized region close to the flock front in the high $\phi_{\rm o}$ regime.} Therefore, rather than describing the obstacles as point-wise scatterers, we here consider small spatial fluctuations  around a homogeneous obstacle density field. The resulting hydrodynamic equations, derived in a Supplementary Note, are  analogous to the Navier-Stokes equations for a polar active fluid:
\begin{align}
\partial_t\rho+\nabla\cdot(\rho\mathbf \Pi)&=0,\label{Massconservation}\\
\partial_t \mathbf \Pi+v_0\mathbf \Pi\cdot\nabla\mathbf \Pi=\mathbb P\cdot&\left [-\beta\nabla\rho+\alpha_2\Delta(\rho \mathbf\Pi)\right.\label{Eqhydropolairedesorde}
\\
&+\gamma\tilde\Delta\cdot(\rho \mathbf \Pi)
+\left.\mathbf F_{\rm o}\right],\nonumber
\end{align}
where we introduce the local polarization $\mathbf \Pi(\mathbf r,t)\equiv\mathbf J(\mathbf r,t)/[v_0\rho(\mathbf r,t)]$, and $\mathbb P=\mathbb I-\mathbf \Pi\mathbf\Pi$. The convective term on the l.h.s.\ of Eq.~\ref{Eqhydropolairedesorde} stems from self-propulsion, $\beta\nabla \rho$ is a pressure term due to the repulsive interactions between the rollers, and 
$\alpha_2$ and $\gamma$ are the elastic constants of this polar liquid ($\tilde\Delta$ is an anisotropic second-order operator). {\color{bleu}Finally disorder is captured by the quenched force field $\mathbf F_{\rm o}(\mathbf r)=-\beta_0\mathbf \nabla\phi_{\rm o}(\mathbf r)$ which focalizes  the rollers  in the valleys of an effective potential given by  the local obstacle-density field $\phi_{\rm o}(\mathbf r)$. }
The linear response of  $\mathbf \Pi=\hat{\mathbf x}+\delta\theta\hat{\mathbf y}$ provides a physical insight into the formation of sparse flowing channels. Within this approximation, the orientational  fluctuations  are readily computed from Eqs.~\eqref{Massconservation} and ~\eqref{Eqhydropolairedesorde}: 
\begin{equation}
\overline{|\delta\theta_{\mathbf q}|^2}=\frac{\beta_{\rm o}^2\phi_{\rm o}}{v_{\rm o}^2+q_x^2(\alpha_2-\gamma)^2\rho^2}\left(\frac{q_y}{q_x}\right)^2
\label{fluctuations}
\end{equation}
where $\overline{\, \cdot\,} $ stands for average over disorder, and $\mathbf q$ is a quasi longitudinal wave vector, see Supplementary Note. 
Eq.~\eqref{fluctuations} establishes that the polarization  fluctuations  are set by  the competition between random stirring, self-propulsion and bending elasticity.  
Importantly,  orientational fluctuations increase at all scales with the number of obstacles. However, the bending stiffness $(\alpha_2-\gamma)$ suppresses the small wavelength fluctuations required to explore the valleys of  $\phi_{\rm o}(\mathbf r)$ which become increasingly branched and curved as the number of obstacles increases. This competition therefore selects a small subset of all the possible paths and consistently  accounts for the formation of  sparse and tortuous river networks upon increasing  disorder. {\color{bleu} This scenario is further confirmed by experiments performed in periodic  lattices of obstacles. By construction, these arrangements display minute density  fluctuations.  Therefore, the random stirring force in Eq.~\eqref{Eqhydropolairedesorde} is  expected to be vanishingly small compared to an equally dense disordered medium. In agreement with our prediction, we find that no river network emerges  in periodic lattices, see Supplementary Figure S7 and  Supplementary Video 5.}
This  scenario is expected to qualitatively hold beyond linear response. In addition, as it does not depend on the specifics of the colloidal rollers it must be relevant to any flock made of  motile bodies obstructed by repelling obstacles, from living-creature groups to swarming robots to soft-active materials.


\section*{Methods}
We use fluorescent Polystyrene colloids of diameter $4.8\,\rm \mu m$ dispersed in a 0.15 mol.L$^{-1}$ AOT-hexadecane solution (Thermo scientific G0500). The suspension is injected in a wide microfluidic chamber made of two parallel  glass slides coated by a conducting layer of Indium Tin Oxyde (ITO) (Solems, ITOSOL30, thickness: 80 nm)~\cite{Bricard2013}. The two electrodes are assembled with double-sided scotch tape of homogeneous thickness ($110\,\mu \rm m$). 
The colloids are confined in a $1\,\rm mm\times 15\, mm$ channel, by walls  made of a positive photoresist resin (Microposit S1818, thickness: 2 $\mu$m). Identical cylindrical obstacles of radius $5\,\mu\rm m$ made of the same material are included in the main channel. Their position is uniformly distributed with a density $\rho_{\rm o}$, and the obstacle fraction is defined as $\phi_{\rm o}=\pi a^2 \rho_{\rm o}$ in the main text. Note that some of the obstacles  overlap. This geometry is achieved by means of conventional UV lithography. More details about the design of the microfluidic device are provided in the  Supplementary Methods.  
 
 The Quincke electro-rotation of the colloids is controlled by applying a homogeneous electric field transverse to the two electrodes $\mathbf E=E_0\hat{\mathbf z}$. The field is applied with a voltage amplifier (TREK 609E-6). All the reported results correspond to an electric field $E_0=2E_{\rm Q}$, where $E_{\rm Q}$ is the Quincke electro-rotation threshold $E_{\rm Q}=1\, \rm V/\mu m$. All the measurements are performed when a steady state is reached for all the observables. 
 The colloids are observed with a 7.2X magnification with a fluorescent Nikon AZ100 microscope. The movies are recorded with a CMOS camera (Basler ACE) at frame rates of 380 fps.  The particles are detected to sub-pixel accuracy, and the particle trajectories and velocities are reconstructed using  the Crocker and Grier  algorithm \cite{Grier}.  Measurements are performed in a $1.22\, \rm mm\times 0.75\,\rm mm$ observation window. {\color{bleu} All measurements have been systematically repeated for 15 to 18 different flocks crossing the same field of view (different initial conditions). In addition we have used four different realization of the disordered arrangements of obstacles. The observation window was set close to the midpoint of the main channel where all the morphological quantities have reached their stationary values. Measurements performed further away from the walls yield identical results.}

All the colloids roll at constant speed $v_0=1.4\pm0.1\,{\rm mm\ s^{-1}}$. When isolated, their direction of motion freely diffuses on the unit circle with a diffusivity $D=1.6\pm0.1\,{\rm s}^{-1}$. $D$ is defined as the exponential decorrelation rate of the velocity orientation in a  isotropic phase, $\langle {\hat{\mathbf v}}_i(0)\cdot\hat{\mathbf v}_i(t)\rangle_i\sim \exp(-Dt)$, where $\hat{\mathbf v}_i$ is the velocity orientation of the i$^{\rm th}$ roller.

\begin{figure*}
\begin{center}
\includegraphics[width=\textwidth]{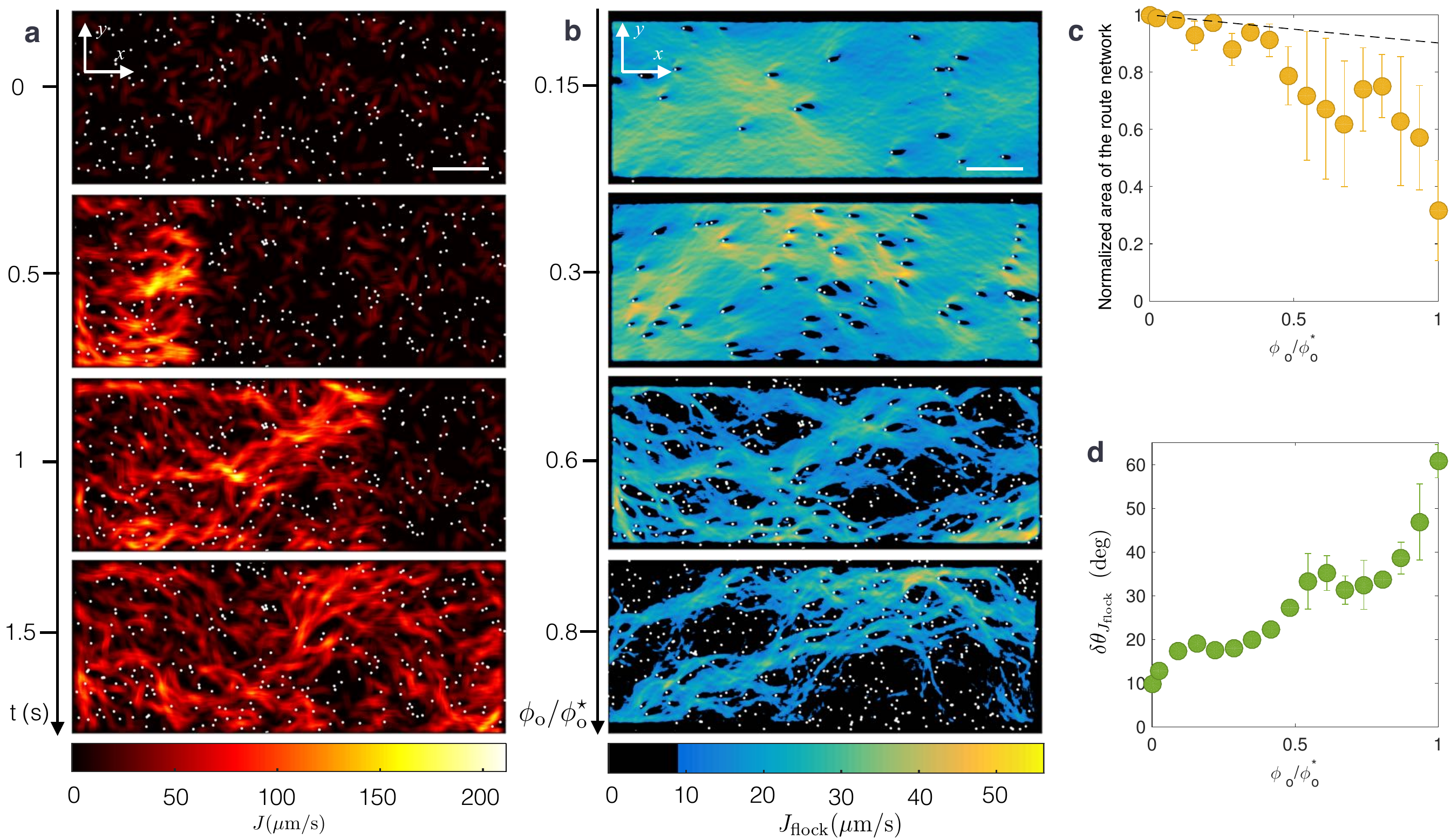}
\caption{{\bf Flocking through river networks.}
\textbf a, Four subsequent snapshots of  the magnitude $|\textbf J(x,y,t)|$ of the current field. Coarse graining over $25\,\rm \mu m\times25\,\mu m$ bins. Time average over $0.05\,\rm s$ $\phi_{\rm o}/\phi_{\rm o}^\star=0.5$. The rollers  flow along preferred channels. Scale bar: $200\,\rm \mu m$. 
\textbf b, Magnitude of  the flock current $J_{\rm flock}$  plotted for four different obstacle packing fractions. At low $\phi_{\rm o}$ colloid-depleted wakes form downstream the obstacles. As $\phi_{\rm o}$ increases, a channel network forms and becomes increasingly sparse and tortuous. Coarse graining over $12.5\,\rm \mu m\times12.5\,\mu m$ bins. Scale bar: $200\,\rm \mu m$. 
\textbf c, Circles: Area of the flowing region  normalized by the area of the observation window. Error bars:  1 sd (17 flocks per data point). Dashed line: area   fraction of  the region left around the  superposition of spatially uncorrelated wakes. This quantity is computed knowing the fraction of space occupied by a random ensemble of patches having the shape of the wake formed downstream an isolated obstacle: $1-\exp\left(-\rho_{\rm o}a_{\rm wake}\right)$, where $\rho_{\rm o}$ is the obstacle density and $a_{\rm wake}$ the wake area~\cite{formulaire}. 
\textbf d, The orientational fluctuations of $\mathbf J_{\rm flock}$ sharply increase at the onset of  flock destruction. They are defined as $\delta \theta_{J_{\rm flock}}^2=A^{-1}\int\! \theta_{J_{\rm flock}}^2(\mathbf r)\,{\rm d}\mathbf r$, where $A$ is the area of the observation region, and $\mathbf J_{\rm flock}/J_{\rm flock}\equiv  \left(\cos\theta_{\rm J_{flock}},\sin \theta_{\rm J_{flock}}\right)$. Error bars:  1 sd (17 flocks per data point).
\label{fig3}}
\end{center}
\end{figure*}

\begin{figure*}[h]
\begin{center}
\includegraphics[width=\columnwidth]{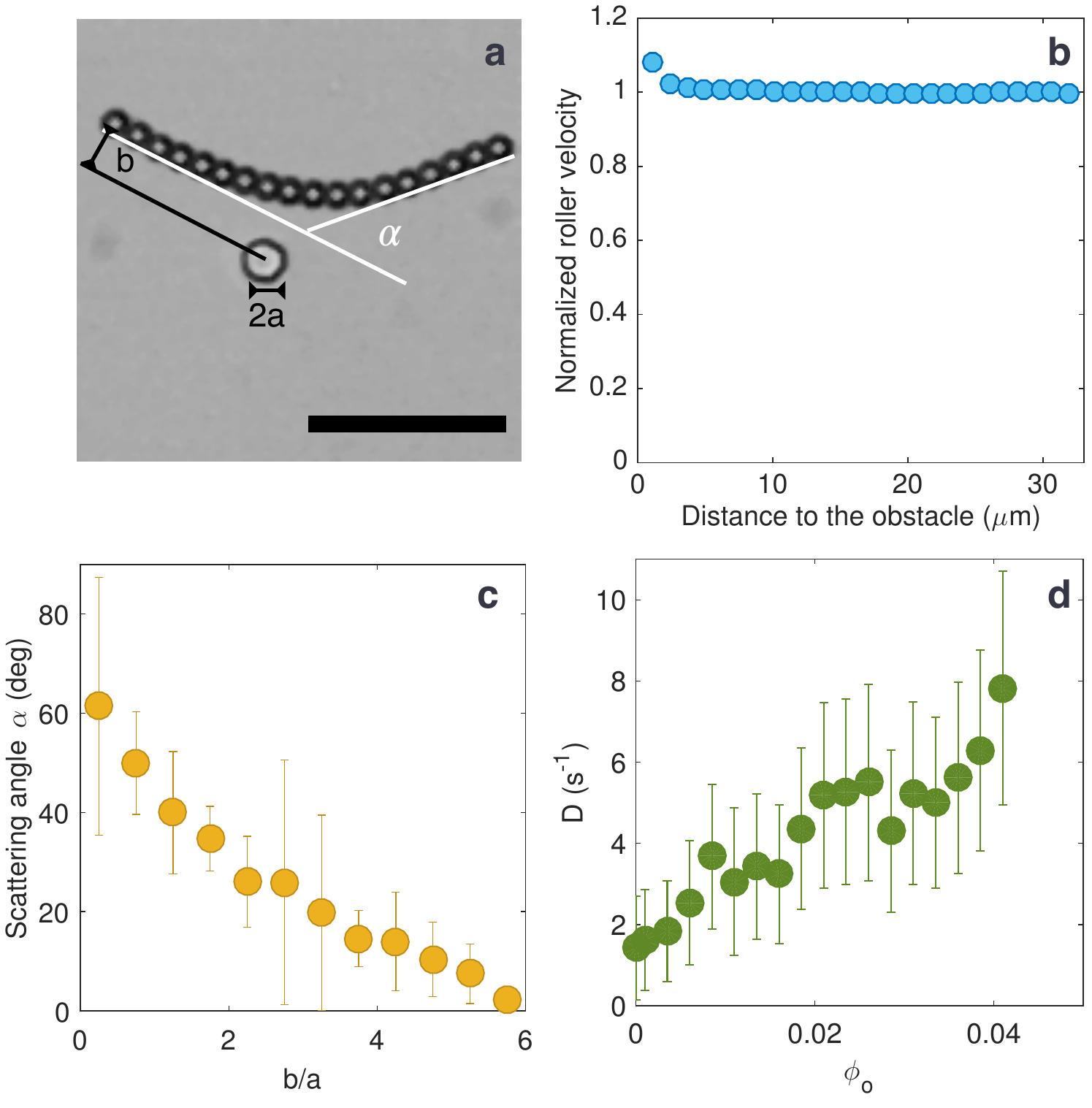}
\caption{{\bf Roller-obstacle scattering.}
\textbf a, Superimposed picture of a roller colliding an obstacle of radius $a=5\,\rm \mu m$. $b$ is the impact parameter, $\alpha$ is the scattering angle. Time interval between each picture: 6.7 ms. $E_{\rm 0}/E_{\rm Q}=1.8$. Scale bar: 50 $\mu$m. \textbf b, Roller velocity normalized by $v_0$ as a function of the distance to the obstacle. Same parameters as in \textbf a. \textbf c,  Scattering angle $\alpha$ plotted versus the normalized impact parameter $b/a$ defined in \textbf a. $E_{\rm 0}/E_{\rm Q}=2$. Error bar: 1 sd. \textbf d, Rotational diffusivity defined as the decorrelation time of the roller velocity plotted as a function of the obstacle packing fraction, see Methods. $E_{\rm 0}/E_{\rm Q}=2$.
\label{fig4}}
\end{center}
\end{figure*}

The current field  $\mathbf J(\mathbf r,t)$ is computed by summing the instantaneous roller velocities in $12.5\,\rm\mu m\times12.5\,\rm\mu m$ binning windows. The flock current $\mathbf J_{\rm flock}(\mathbf r)$ is computed by averaging $\mathbf J(\mathbf r,t)$ over time. The flowing-path network is defined as the ensemble of points where $J_{\rm flock}$ exceeds $11\,\mu \rm m/s$. This value has been chosen as the typical average current in the wake left behind an isolated obstacle. None of the results discussed in this letter qualitatively depends on this specific threshold value. The current-free regions referred to in the main text are associated with local current value  smaller than this threshold  (black  areas in Fig.~\ref{fig3}b).

\noindent {\bf Acknowledgements.} We acknowledge support from ANR program MiTra and Institut Universitaire de France. We thank  F. Peruani, S. Santucci, M. C. Marchetti and  D. Carpentier for valuable comments and suggestions. We also thank G. Fabre for help with the experiments.
 
\noindent{\bf Author Contributions} D.~B. conceived the project.   N.~D., A.~M. and D.~B. designed the experiments. N.~D. and A.~M. performed the experiments. J.-B.~C. and D.~B. performed the theory. N.~D., A.~M., J.-B.~C. and D.~B. analyzed and discussed results. D.~B. and A.~M. wrote the paper.  N.~D. and A.~M. have equal contributions.
 
\noindent{\bf Author Information} Correspondence and requests for materials
should be addressed to D.~B. (email: denis.bartolo@ens-lyon.fr).


\end{document}